\documentclass[twocolumn,amsmath,amssymb]{revtex4-1}

\usepackage{graphicx}
\usepackage{dcolumn}
\usepackage{bm}
\usepackage{color}
\usepackage{soul}

\usepackage[normalem]{ulem}

\newcommand{\dt}{\ensuremath{\widetilde{S}}}
\newcommand{\ig}{\ensuremath{I_{\rm G}}}
\newcommand{\ioff}{\ensuremath{I_{\rm off}}}
\newcommand{\ih}{\ensuremath{I_{\rm H}}}
\newcommand{\vg}{\ensuremath{V_{\rm g}}}
\newcommand{\voff}{\ensuremath{V_{\rm off}}}

\begin{document}


\title{Differential thermopower images as a probe of many-body effects in quantum point contacts}

\author{Yuan Ren}
\author{Joshua A. Folk}
\thanks{to whom correspondence should be addressed: {\it jfolk@physics.ubc.ca}}
\affiliation{Department of Physics and Astronomy, University of British Columbia, Vancouver, BC V6T
1Z1, Canada}
\affiliation{Stewart Blusson Quantum Matter Institute, University of British Columbia, Vancouver, BC V6T
1Z4, Canada}

\author{Yigal Meir}
\affiliation{Department of Physics and the Ilse Katz Institute of Nanotechnology, Ben-Gurion University, Beer Sheva 84105, Israel}
\author{Toma\v{z} Rejec}
\affiliation{Faculty of Mathematics and Physics, University of Ljubljana, Jadranska 19, SI-1000 Ljubljana, Slovenia}
\affiliation{Jo\v{z}ef Stefan Institute, Jamova 39, SI-1000 Ljubljana, Slovenia}
\author{Werner Wegscheider}
\affiliation{Department of Physics, ETH Zurich, CH-8093 Zurich, Switzerland}
\affiliation{Quantum Center, ETH Zurich, CH-8093 Zurich, Switzerland}



\date{\today}





\maketitle


{\bf 
Mesoscopic circuit elements such as quantum dots and quantum point contacts (QPCs) offer a uniquely controllable platform for engineering complex quantum devices, whose tunability makes them ideal for generating and investigating interacting quantum systems.  However, the conductance measurements commonly employed in mesoscopics experiments are poorly suited to discerning correlated phenomena from those of single-particle origin. Here, we introduce non-equilibrium thermopower measurements as a novel approach to probing the local density of states (LDOS), offering an energy-resolved readout of many-body effects. We combine differential thermopower measurements with non-equilibrium density functional theory (DFT) to both confirm the presence of a localized state at the saddle point of a QPC and reveal secondary states that emerge wherever the reservoir chemical potential intersects the gate-induced potential barrier. These experiments establish differential thermopower imaging as a robust and general approach to exploring quantum many-body effects in mesoscopic circuits.

}

Thermal measurements are currently emerging as powerful probes for the investigation of strongly correlated quantum states in low-dimensional systems. For instance, thermal conductance provides strong evidence for non-abelian quasiparticles in fractional quantum Hall samples \cite{banerjee2018observation}, and chemical-potential measurements have demonstrated the emergence of Pomeranchuk-like local moments in twisted bilayer graphene \cite{Rozen:2021va,Saito:2021vu}. With a view to experimental implementations, thermopower is of particular interest, as it can be read out as easily as conductance while still probing thermodynamic characteristics. Thermopower has been proposed as a metric to characterize the non-Fermi liquid nature of strongly interacting quantum states in, for example, multi-channel charge Kondo circuits \cite{PhysRevLett.125.026801} and condensed matter realizations of the SYK model \cite{PhysRevB.101.205148}. However, experimental investigations of mesoscopic thermodynamics are still in their infancy.  Typically, comparisons are made to single-particle predictions of the Mott formula \cite{CutlerMott} as a way of testing for the presence of interactions, but without offering microscopic insight into what role those interactions might play.  Here, we show that non-equilibrium thermopower measured in a mesoscopic system offers an energy-resolved readout of many-body effects in the density of states, that can be compared directly with state-of-the-art modelling.

As an initial test for this approach, we turn our attention to the most prototypical of mesoscopic devices, the quantum point contact (QPC).
QPCs are short saddle-point constrictions in a 2D sheet of high-mobility electrons whose transmission can be tuned by a gate voltage \cite{WeesPRL88, WharamJPC88, ButtikerPRB90}.  Near pinchoff, interactions dominate the potential landscape seen by electrons in a QPC, and it has been proposed that localized states emerge in and near the constriction due to interactions \cite{Hirose:2003bt,rejecNature06,  ZozoulenkoPRB09,PhysRevB.77.041301,PhysRevB.77.085314}.  Such localized states can, for example, explain the phenomenology of an extra plateau --- the famed 0.7 structure --- that appears in the conductance of the cleanest QPCs \cite{ThomasPRB98,CronenwettPRL02}.
Beyond their ability to explain the conductance characteristics of 0.7 structure, however, direct evidence for localized states remains scarce \cite{Iqbal2013,Brun2014}, and what evidence there is leaves open the possibility of different interpretations \cite{Bauer13}. Here, we demonstrate that non-equilibrium thermopower data offers a complete spectroscopic map of the energy of localized states in a QPC 
and their effect on the local density of states, enabling a direct comparison to DFT (or other) models as a test of their accuracy.

\begin{figure}
\includegraphics{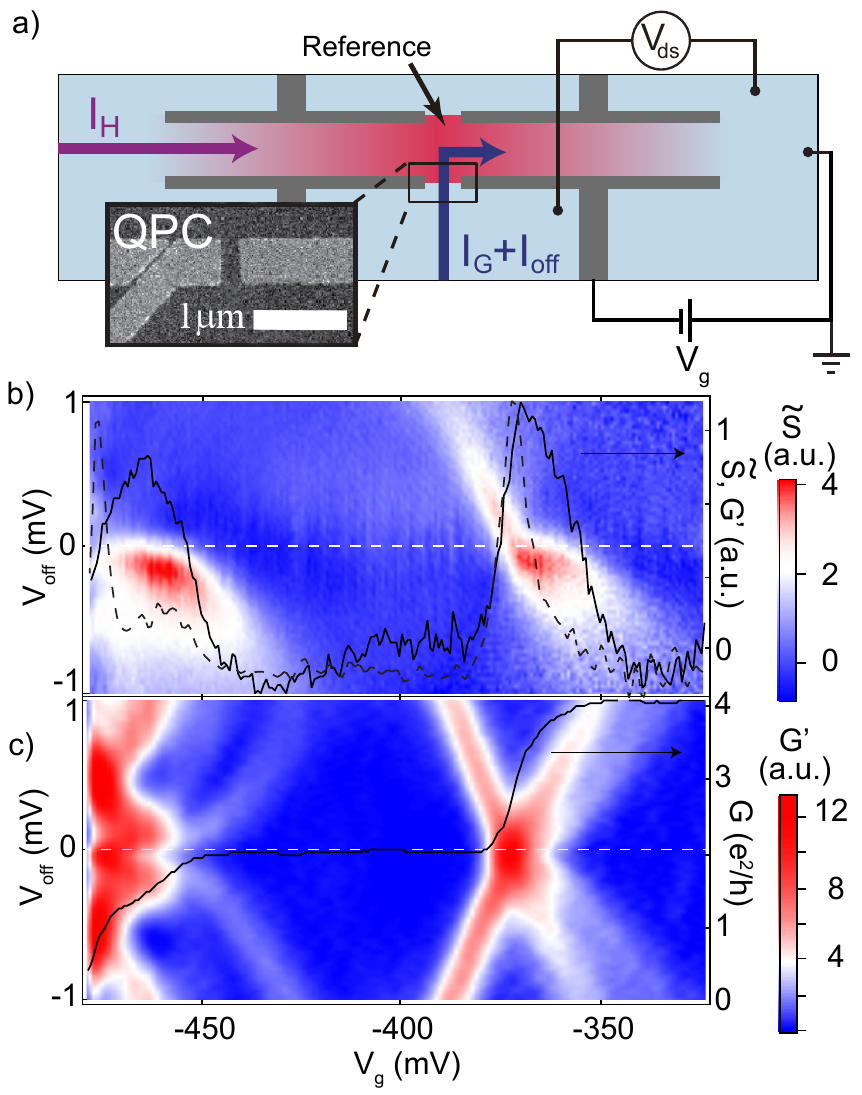}
\caption{\label{fig-intro}  ({\bf a}) Device geometry and measurement setup showing currents for heating (\ih), monitoring QPC conductance (\ig) and offsetting  chemical potentials of hot and cold reservoirs (\ioff), and the $V_{\rm ds}$ measurement.  Light blue: cold 2DEG; red:  hot 2DEG; grey: electrostatic gates.  Reference QPC is discussed in Methods.  Inset: scanning electron micrograph shows QPC geometry.  ({\bf b}) Differential thermopower  $\dt\equiv S\cdot G$ over the first two conductance steps for a typical QPC (QPC1 shown here); line plots for $\voff=0$ (right axis) show peaks in $\dt$ (solid) as well as $G'$ (dashed) for each step in $G$.  ({\bf c})   Transconductance ($G'\equiv dG/d\vg$) for QPC1;   Colorscale data plotted as a function of $\vg$ (bottom axis) and $\voff$ (left axis).    Line plot (right axis) of conductance $G$ for $\voff=0$ shows characteristic steps in $G$ and clear 0.7 structure.}
\end{figure}

Thermopower, $S$, quantifies the excess voltage that appears across a device due to a temperature difference, $T_{\rm ds}\equiv T_{\rm d}-T_{\rm s}$, between drain and source reservoirs when the current through the device  fixed: $S=-\partial V_{\rm ds}/\partial T_{\rm ds}$.
$S$ typically reflects an imbalance between the electron flow above the chemical potential, $\mu$, and hole flow below $\mu$. That imbalance yields a temperature-driven current that must be counteracted by a voltage to keep the net current constant.  Multiplying by the differential conductance, $G$, the product $\dt\equiv S\cdot G$ is often more useful to analyze than $S$ itself as it provides direct access to the temperature-driven current rather than the voltage that builds up to oppose it.  \dt\ is connected to the derivative of the energy-dependent transmission through the device at $\mu$, $dt/d\varepsilon|_{\varepsilon=\mu}$, making it sensitive to fine features in the density of states and offering therefore an important advantage over conductance, which depends on $t$.  Early measurements of QPC thermopower confirmed peaks in \dt\ at the entry of each new subband, where $dt/d\varepsilon$ is large \cite{Kozub:1994ga, Houten:1999kj}, whereas later experiments identified significant deviations from the thermopower that would be expected in a simple picture of non-interacting 1D subbands, and inspired the measurements presented here \cite{Appleyard:2000iz}.

Although \dt\ is often measured with the current fixed at zero, differential measurements of \dt\ may also be performed with a bias current that offsets source and drain chemical potentials by $e\voff\equiv \mu_{\rm d}-\mu_{\rm s}$ (\voff\ is defined in the limit  $T_{\rm ds}=0$).  The temperature of one of the reservoirs is then oscillated to measure $\partial V_{\rm ds}/\partial T_{\rm ds}$.
In mesoscopic systems such as  QPCs, \dt(\voff) offers a spectroscopic readout of thermopower that quantifies $dt/d\varepsilon$ at the chemical potential of the reservoir whose temperature changes \cite{Appleyard:2000iz}.  This represents a further distinct advantage of \dt\ over standard measurements of differential conductance at finite bias, which superimpose features at both source and drain potentials. (See, for example, Fig.~1b vs 1c.)

In this paper, we analyze $\dt(\voff)$ spectra of QPCs covering weakly to strongly interacting regimes [Fig.~1b].  The data are broadly consistent with published experiments \cite{Appleyard:2000iz}, but a careful examination reveals robust signatures that are inconsistent with existing theoretical models.  The new realization described in this paper is that those signatures are in fact a direct readout of many-body modifications to the density of states in the QPC.
In order to clarify this connection, we extend density functional theory (DFT) previously developed to describe the microscopic state of QPCs at equilibrium, 
to incorporate non-equilibrium effects on the potential due to finite bias and different source/drain temperatures. Although exact steady-state DFT has, in principle, been formulated for non-equilibrium conditions \cite{Stefanucci15,Sobrino19}, a Landauer+DFT approach \cite{PhysRevB.51.5278,PhysRevB.52.5335}
is used here (see the SI) due to the lack of a corresponding exchange-correlation functional.

The result of our analysis is a more complete understanding of QPCs as strongly interacting mesoscopic elements: the convenient but naive picture in which gate voltages uniquely define a saddle-point potential barrier evolves to one in which the self-consistent barrier that determines transmission is additionally a result of charge and spin degrees of freedom at the barrier responding to non-equilibrium conditions imposed by the reservoirs. 
Our data highlight the critical role of localized states that emerge when the bottom of a 1D subband crosses the chemical potential.  These states were previously invoked to explain 0.7 structure in QPC conductance
\cite{Meir2002,MeirJPCM08}; here they explain striking thermopower signatures that defy other explanation.

QPCs were defined in the wall of a micron-wide channel (Fig.~1a). The interior of the channel formed the grounded drain of the QPC, which could be Joule-heated to $T_{\rm d}\sim150\,$mK by passing current \ih\, while the source remained at base temperature ($T_{\rm s}=100\,$mK) due to its low-resistance connection to an ohmic contact.  Independent AC currents $\ih$ and $\ig$ enabled simultaneous lock-in measurement of $S$ and $G$, while $\ioff$ at DC was used to offset $\mu_{\rm s}$ and $\mu_{\rm d}$, see Methods.
The thermal voltage across the QPC was dominated by diffusion thermopower, with phonons nearly absent at $100\,$mK.

Figure~1b shows \dt\ for a typical QPC.  For comparison, Fig.~1c shows the trans-conductance, $G'\equiv dG/d\vg \propto dG/d\mu$, another metric that is related to $dt/d\varepsilon$ when $\vg$ controls $\mu$.  This connection is, however, indirect in interacting systems where $t(\varepsilon)$ itself depends on $\mu$, and when the shape of the potential changes with $\vg$ as is often the case in mesoscopic devices.  Because $G'$ is based on conductance, Fig.~1c is symmetric around $\voff=0$, superimposing features at the source chemical potential (lower left to upper right in the image) on others at the drain (upper left to lower right).  For the discussion that follows, the lack of symmetry in $\dt$ [Fig.~1b] is greatly clarifying, distinguishing responses at negative energy from those at positive energy (below vs above the average $\mu$).  This will, for example be key to identifying the role of localized states in the analysis to come.

\begin{figure}
\includegraphics[scale=1.2]{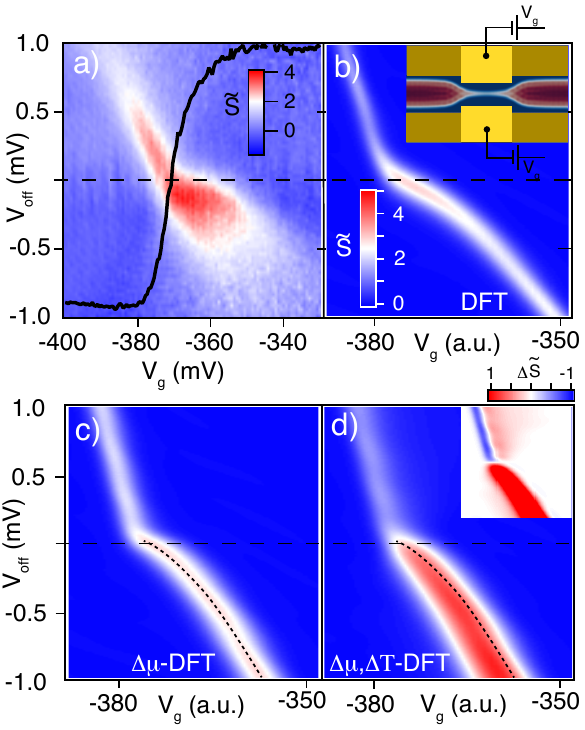}
\caption{\label{fig-intro}  (a-d) Evolution of thermopower ($\dt$, colorscale) with $\voff$  across the entry of the second subband into the QPC.  Experimental data from QPC1 (a) is compared to simulations (b,c,d) of $\dt$ calculated using the screened potential from local gates. The potential in (b) [DFT] is calculated at $T_{\rm s}=T_{\rm d}=100\,$mK and in equilibrium; in (c) [$\Delta \mu$-DFT], at $T_{\rm s}=T_{\rm d}=100\,$mK but considering different source/drain potentials; and in (d) [$\Delta \mu$,$\Delta T$-DFT], considering different source/drain potentials as well as different temperatures. Dotted line in (c) traces the maximum of $\dt$; the identical line brought to (d) illustrates the spreading of large $\dt$ to the left due to temperature-dependent screening.  (d, inset) shows the numerical difference between $\dt$ from (c) and (d), main panels.}
\end{figure}

We first consider measurements in a regime of relatively weak interactions, at the $2e^2/h\rightarrow 4e^2/h$ transition where the second subband drops below the chemical potential.  Figure~2 compares experimental data to three levels of numerical simulation.  Fig.~2b [DFT] represents an equilibrium calculation: the QPC potential and LDOS are calculated self-consistently for each $\vg$, assuming $\voff=0$ and $T_{\rm s}=T_{\rm d}=100\,$mK, with $t(\varepsilon)$ obtained from Kohn--Sham scattering states\cite{PhysRevB.51.5278,PhysRevB.52.5335}. $\dt(\voff)$ is  determined from the simulation by comparing the current through the device for \{$T_{\rm d}=150\,$mK, $T_{\rm s}=100\,$mK\} and $T_{\rm d}=T_{\rm s}=100\,$mK, where current is calculated by integrating $t(\varepsilon)$ over thermally broadened source and drain reservoirs offset by $\voff$.

Figures~2c [$\Delta\mu$-DFT] and 2d [$\Delta\mu$,$\Delta T$-DFT] bring in the non-equilibrium character of the experiment at increasing levels. First, in Fig.~2c [$\Delta\mu$-DFT] the LDOS and $t(\varepsilon)$ are recalculated for each \{$\vg$, $\voff$\} pair, including different reservoir potentials but still assuming $T_{\rm s}=T_{\rm d}=100\,$mK.  As before, \dt\ is determined for each $t(\varepsilon)$ by comparing the currents at $T_{\rm d}=150$ and $100\,$mK. 
Fig.~2d [$\Delta\mu$,$\Delta T$-DFT] extends this to include different values of $T_{\rm d}$ from the beginning, that is, including the effect of the elevated drain temperature on the LDOS and thereby on $t(\varepsilon)$.

All simulations capture the diagonal motion of the subband, with a kink in the slope at $V_{\rm off}=0$ due to the divergence in the 1D density of states at the bottom of of the subband.  But Figs.~2b,c miss the broadening of the subband thermopower after crossing $\voff=0$, which only appears once finite temperature difference is taken into account (Fig.~2d). The difference between Fig.~2c and 2d, shown in the inset of Fig.~2d, highlights one of the most important insights to be gained from this work: temperature affects mesoscopic transport, including thermopower, not only by broadening the reservoir Fermi distributions (Figs.~2b,c), but also by changing the screened QPC potential itself, the LDOS, and thereby $t(\varepsilon)$  (Fig.~2d).
We take advantage of this sensitivity to probe sharp features in the LDOS, which are more strongly affected by a change in temperature.

For insight into this effect, consider the full electrostatic potential seen by an incoming electron just at the entry of the second subband, including the potential  defined by electrostatic gates as well as repulsion from electrons occupying the first or second subbands.  With $\mu$ at the bottom of second subband, the density of states is significantly larger just above $\mu$ than below, so  thermal broadening around $\mu$ leads to a net increase in electron density (more electrons occupying the second subband).
This extra negative charge creates a more repulsive potential for the incoming electron, reducing the total (positive) current and suppressing $\dt$ when $\voff>0$.
The contribution is opposite when the potentials are reversed, yielding the broad positive feature in $\dt$ when $\voff<0$.  For further discussion of Fig.~2d, see the SI.

\begin{figure*}
\includegraphics{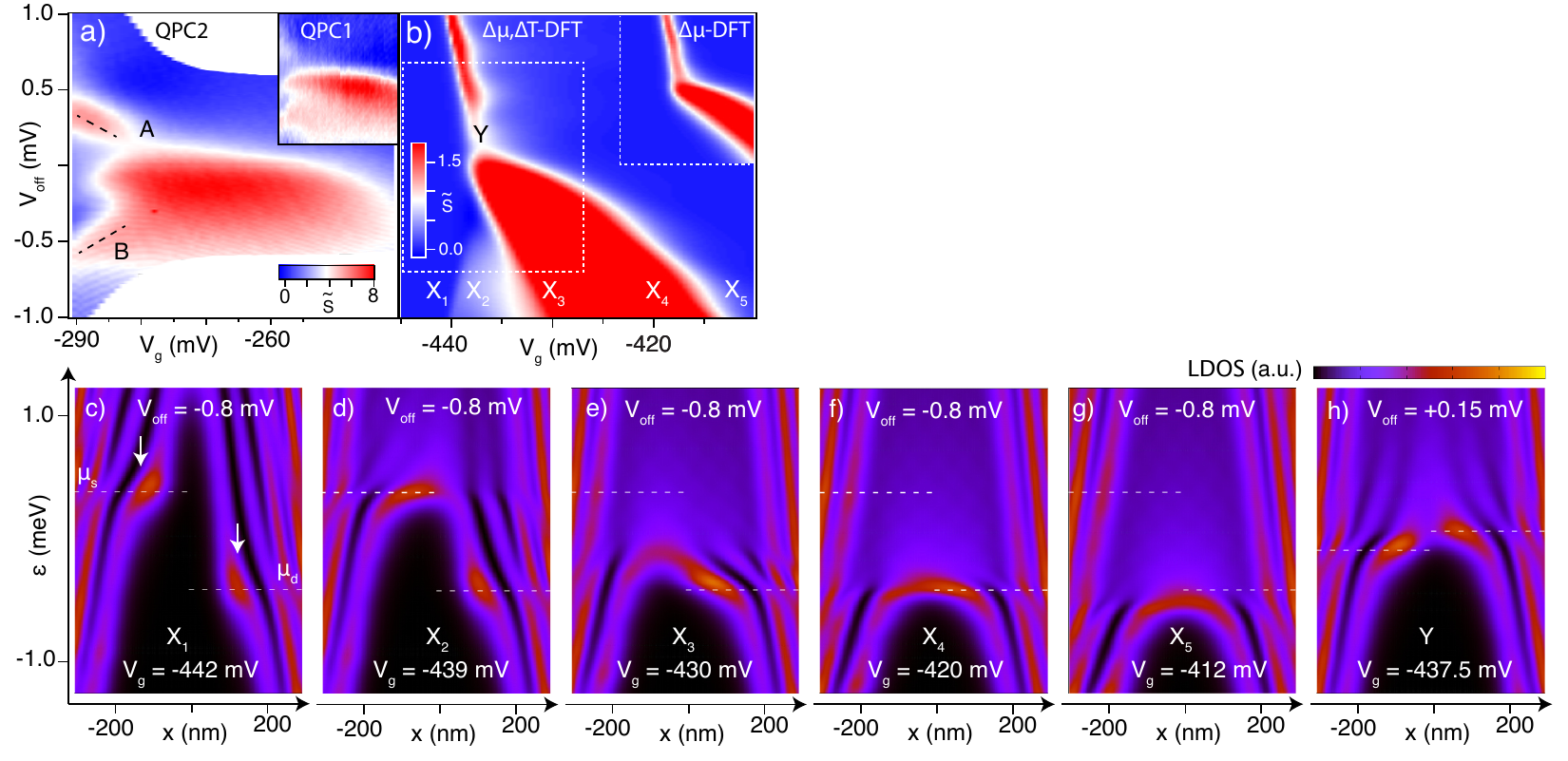}
\caption{\label{fig-intro}  (a) Characteristic signatures at the entry of the first subband include a significant weakening of the subband thermopower just before it crosses the chemical potential (`A') and a broad region of strong thermopower at negative \voff, extending into a diagonally moving feature deep in the pinchoff regime (`B'). Main panel data are from QPC2; inset shows equivalent data from QPC1.  (b) The $\Delta\mu$,$\Delta T$-DFT simulation reflects all signatures of the data (saturated colorscale); inset shows the $\Delta\mu$-DFT simulation over region surrounded by dotted line in main panel.  (c-g) LDOS along the length of the QPC (horizontal axis), versus energy (vertical axis), for $\voff=-0.8\,$mV and gate voltages marked `X$_i$' in panel (b).  Dashed lines show $\mu_{\rm s}=0.4\,$meV and $\mu_{\rm d}=-0.4\,$meV.  Localized states are indicated by arrows. (h) Equivalent LDOS map for point `Y' in panel (b).}
\end{figure*}

Building off this insight, we turn to the entry of the first subband into the QPC, the $0\rightarrow2e^2/h$ transition, where interactions are widely believed to dominate conductance phenomenology (Fig.~3a). 
As in the case of the second-subband data, the feature moving diagonally down from the upper left reflects the subband lowering in energy with more positive gate voltage. In contrast to Fig.~2a, however, this feature reduces significantly in strength just before crossing $\voff=0$ (location `A' in Fig.~3a), then after crossing $\voff=0$ it broadens and strengthens much more dramatically than for the second subband.  In fact, the extension of this feature to more negative $\vg$ is so extreme as to create an extra `leg' of strong $\dt$ that moves down and to the left, as if associated with the movement of the source instead of the drain (`B' in Fig.~3a).  Analogous features were observed in all QPCs measured.

These new features in the first-subband thermopower are clearly reproduced in the $\Delta \mu,\Delta T$-DFT (Fig.~3b), but are missing from simulations that leave temperature difference out of the self-consistent potential calculation (inset in Fig.~3b). This demonstrates that they originate in the temperature-dependent screening effect described previously.  In order to understand this phenomenology more fully, we examine the LDOS at several points in Fig.~3b.  Consider, first, the row of points X$_1\rightarrow$X$_5$ at $\voff=-0.8\,$mV, for which energy-dependent LDOS maps along the axis of the QPC are illustrated in Fig.~3c--g.  Localized enhancements of the LDOS (localized states) are clearly visible anywhere $\mu_{\rm s}$ or $\mu_{\rm d}$ intersect the potential barrier, especially noticeable in Fig.~3c--e.  The fact that these states are pinned close to $\mu_{\rm s}$ or $\mu_{\rm d}$ is key to their strong effect on the thermopower.

The simplest cases are the extremes, X$_1$ and X$_5$.  At X$_1$, $\dt$ is zero simply because the barrier is well above $\mu_{\rm s}$ and $\mu_{\rm d}$, so that no current can flow ($dt/d\varepsilon=0$ as $t=0$ at both $\mu_{\rm s}$ and $\mu_{\rm d}$).  At $X_5$, $\dt=0$ again because $dt/d\varepsilon=0$ in the fully transmitting first subband, where both chemical potentials lie.  Next simplest is X$_4$, where large $\dt$ is expected from a conventional thermopower picture, due to the alignment of $\mu_{\rm d}$ with the bottom of the subband, where $dt/d\varepsilon$ is maximum.

Points X$_2$ and X$_3$ are more interesting. At X$_2$, for example, the conventional picture would predict  $\dt=0$, because  $t$ and  $dt/d\varepsilon$ are both zero at $\mu_{\rm d}$, and  the large $dt/d\varepsilon$ at $\mu_{\rm s}$ does not contribute to the thermopower because $T_{\rm s}$ is not varied. However, significant $\dt$ is observed, in both  simulation and experiment.    A closer analysis of the simulation results shows that $\dt$ appearing at this location is due to temperature-induced charging of the localized state at $\mu_{\rm d}$, which affects the barrier potential and thereby the current.  This effect is even stronger at $X_3$, where the localized state is sharper and the barrier maximum is far from either reservoir potential, free to shift as the localized state charges.

Thermopower observed at $X_3$ is analogous to, but much stronger than, the broadening of second-subband thermopower in Fig.~2d, compared to Fig.~2c.  In both cases, temperature-induced charging of the LDOS where $\mu_{\rm d}$ intersects the bottom of the subband shifts the entire subband up in energy, tending to reduce the current.  This broadening is stronger for the first subband due to the localized state at $\mu_{\rm d}$, as seen from the significantly wider extent of large $\dt$ in Fig.~3b compared to Fig.~2d.

The fact that a large $\dt$ remains all the way to $X_2$ in Fig.~3b can be traced, in the simulation, to the second localized state at $\mu_{\rm s}$ (see Fig.~3d).  The exchange term included in the DFT is what allows for the emergence of localized states; analogous simulations without the exchange term retain the thermopower at $X_3$, but the extension to $X_2$ (which maps to feature `B' in the data) is gone.  The localized state at $\mu_{\rm s}$ makes the potential at the barrier top especially sensitive to charging of the state at $\mu_{\rm d}$, just as quantum dots at a Coulomb blockade transition charge sharply in response to a small changes in gate voltage (see the SI).

The weakening of thermopower just above $\voff=0$, marked $Y$ in Fig.~3b, is again a signature that appears only when temperature-induced screening is taken into account, and can be motivated by the same qualitative arguments as above.  As seen in Fig.~3h,  a localized state is pinned at $\mu_{\rm d}$, but in this case $\mu_{\rm d}$ is above $\mu_{\rm s}$. Thermally induced charging of the localized state (which raises the QPC barrier) therefore reduces the current from drain to source, opposing the conventional sign of thermopower and leading to the reduction seen at $Y$.

It is remarkable that this qualitative picture, involving a localized state becoming more occupied (more negatively charged) at higher temperatures, so consistently explains the sign of thermopower modifications throughout the pinchoff region.  These modifications depend on the derivative of 1D subband LDOS with respect to energy: when $\mu_{\rm d}$ is just below a peak in the LDOS, the state will charge with $T$; when it is just above, the state will discharge, and the opposite sign of thermopower will be observed.  The consistency of the sign observed in both experiment and simulation is therefore a clear indication that the localized state is pinned to just above $\mu$, regardless of the details of the QPC electrostatic potential.

This pinning of the localized state just above $\mu$ was a characteristic specifically of the unpolarized solution of the DFT equations, even though there was another, polarized, solution that, at equilibrium, was lower in free energy.
Polarized DFT solutions are ones in which the LDOS may be different between the two spin species; the thermopower calculated from this solution did not match the experiment, see the SI. The unpolarized solution, presented throughout this paper, forces them to remain the same.  The fact that the unpolarized solution more closely matches the data supports the role of Kondo physics in the low-energy spectrum of QPCs:  Kondo correlations would tend to lower the energy of the spin-singlet (unpolarized) state, but these correlations are not captured in the local density functional used here.

In conclusion, we have discovered a new route to probing the LDOS in  mesoscopic structures, by differential thermopower measurements at finite bias.  Whereas conventional thermopower is a single-particle effect that probes energy-dependent scattering, this new mechanism is based on interactions and reflects thermal screening of the self-consistent potential landscape due to an energy-dependent density of states.  As 0.7 structure in QPC conductance is difficult to explain without a localized state, our data reflect robust signatures in QPC thermopower, covering a large fraction of parameter space near pinchoff, that also cannot be explained without this state.  Remarkably, a non-equilibrium DFT calculation that includes mean-field Coulomb interaction, spin via an exchange potential, \emph{and} finite temperatures in the leads, reproduces nearly all features of the thermopower data with no fine-tuning of parameters.

METHODS: Multiple bias currents are used in this experiment to measure $S$ and $G$ independently, and to control the chemical potential offset $\voff$ between source and drain reservoirs.  The measurement of $G$ is analogous to nonlinear conductance data reported by many groups in the past, albeit in a less-conventional current biased configuration.  A small oscillating $\ig$, at frequency $f_{\rm G}$, is added to a DC $\ioff$.  Locking in to oscillations in $V_{\rm ds}$ at $f_{\rm G}$ then yields the differential conductance $G$ at a $V_{\rm off}$ that is fixed by $\ioff$.  Although $V_{\rm off}$ can in principle be measured directly for a given $\ioff$ as the DC component of $V_{\rm ds}$, in practice we find it by a numerical integration of $G^{-1}$ over $\ioff$, an approach that was found to yield less measurement noise while giving consistent values.
 
The measurement of $S$ proceeds in a similar way: temperature is oscillated by Joule heating due to current bias $\ih$ through the channel.  With $\ih$ oscillated at frequency $f_{\rm H}$, the thermal voltage is detected by locking in to $V_{\rm ds}$ at $2f_{\rm H}$ as $\ioff$ is scanned (as above) to investigate $S$ for varying source and drain chemical potentials. The combination of two AC bias currents $\ih$ and $\ig$ and one DC current $\ioff$ thus enables a simultaneous measurement of $S$ and $G$ while scanning (and measuring) $V_{\rm off}$.  $\ih$ on the order of $10\,$nA elevated the temperature by $60\pm20\,$mK, estimated using the magnitude of $S$ for $\ioff=0$ at higher subband transitions, where thermopower in QPCs is well understood. 
 
In practice the measured $V_{\rm ds}$ includes contributions both from QPC thermopower and from the 2DEG in the channel.  The extra 2DEG contribution can be quantified by measuring the thermal voltage across the reference QPC set at on a condutance plateau, $G_{\rm ref}=2e^2/h$ where the thermopower is zero, but in practice this contribution was found to be negligible.

ACKNOWLEDGEMENTS: This project has received funding from European Research Council (ERC) under the European Union’s Horizon 2020 research and innovation programme under grant agreement No 951541. YM acknowledges support by the Israel Science Foundation (grant 3523/2020). TR acknowledges support by Slovenian Research Agency under Contract No. P1-0044.  JF acknowledges support from the Stewart Blusson Quantum Matter Institute, the Natural Sciences and Engineering Research Council of Canada, the Canada Foundation for Innovation, the Canadian Institute for Advanced Research, and the Canada First Research Excellence Fund, Quantum Materials and Future Technologies Program.
\clearpage

\par

\bibliography{2010QPCbib}

\end{document}



\title{Supplementary information: Differential thermopower images as a probe of many-body effects in quantum point contacts}


\author{Yuan Ren}
\author{Joshua A. Folk}
\thanks{to whom correspondence should be addressed: {\it jfolk@physics.ubc.ca}}
\affiliation{Department of Physics and Astronomy, University of British Columbia, Vancouver, BC V6T
1Z1, Canada}
\author{Yigal Meir}
\affiliation{Department of Physics and the Ilse Katz Institute of Nanotechnology, Ben-Gurion University, Beer Sheva 84105, Israel}
\author{Toma\v{z} Rejec}
\affiliation{Faculty of Mathematics and Physics, University of Ljubljana, Jadranska 19, SI-1000 Ljubljana, Slovenia and Jo\v{z}ef Stefan Institute, Jamova 39, SI-1000 Ljubljana, Slovenia}
\author{Werner Wegscheider}
\affiliation{Department of Physics, ETH Zurich, CH-8093 Zurich, Switzerland and Quantum Center, ETH Zurich, CH-8093 Zurich, Switzerland}



\date{\today}





\maketitle

\clearpage

\section*{S1. Landauer+DFT calculations}

We model the QPC as a set of three electrostatically coupled strictly two-dimensional systems: the electrodes, the donor layer and the 2DEG. The structure of the simulation closely models the one used in the experiment, with small modifications in simulated charge density to more closely replicate actual gate voltages used in the experiment. 

There are two sets of surface electrodes. The dark yellow ones in the inset to Fig.~2b of the main article, separated by $300\,$nm and biased to $-285\,$mV, define a wide quantum wire in the 2DEG. The light yellow ones are gate electrodes defining the QPC with a lithographic width and length of $200\,$nm and $500\,$nm, respectively.   

A uniformly doped fully ionized donor layer is $55\,$nm below the surface and provides an electron density of $2.5\times10^{11}\,$cm$^{-2}$.  The 2DEG (brown in the inset to Fig.~2b of the main article) is $110\,$nm below the surface. Away from the QPC, the two-dimensional regions of 2DEG are modeled by wide quantum wires, making the system quasi-one-dimensional with six open channels per wire. Far away from the QPC the width of the 2DEG is $\sim200\,$nm and the electron density at the center of a wire is $\sim10^{11}\,$cm$^{-2}$.

Due to screening, the potentials and the electron densities on surface and in the 2DEG are independent of the distance from the QPC far away from the QPC. Therefore, we separated the system into two semi-infinite uniform quantum wires and a $1600\,$nm long central region, containing the QPC at its center. 

We treated the 2DEG quantum-mechanically within the density functional theory (DFT), using the local spin-density approximation with the Tanatar-Ceperley exchange-correlation functional \cite{Tanatar89}. The presence of the semiconductor was taken into account through the use of the GaAs dielectric constant $\kappa = 12.9$ and the effective mass of 0.067 times the bare electron mass.
 
In each iteration of the DFT self-consistency loop, we first obtained the Kohn-Sham scattering states. We projected the 2D Kohn-Sham Hamiltonian to the lowest 10 transverse modes at each point along the QPC axis and then solved the resulting quasi-1D scattering problem by matching a wavefunction in the central region to asymptotic plane and evanescent waves in the uniform quantum wires.

Using the Kohn-Sham scattering states we calculated the 2DEG density using the Fermi functions with in general different chemical potentials and temperatures for electrons incoming from the source and from the drain electrodes. We applied the bias voltage symmetrically to the source and to the drain.

We then distributed the remaining electrons on the surface in such a way that the electrostatic potential there assumed the required values. For the exposed surface we used the "pinned" surface boundary condition \cite{Davies1995}. In this step we performed the calculation on a wide strip of width $800\,$nm centered at the QPC axis, with periodic boundary conditions, which enabled us to use the fast Fourier transform method. With the resulting charge distribution we calculated a better approximation to the electrostatic potential in the 2DEG to be used in the next iteration of the DFT self-consistency loop. 

\section*{S2. Device fabrication}

The mesoscopic circuits measured in this work were defined using electrostatic CrAu gates on the surface of a GaAs/AlGaAs heterostructure.  The 2DEG was $110\,$nm below the surface, with electron density $n_s = 1.11 \times 10^{11}\,$cm$^{-2}$ and mobility $\mu = 4.44 \times 10^6\,$cm$^2$/Vs measured at $T = 1.5\,$K. The heating channels had a lithographic width $1\,\mu$m and length $100\,\mu$m, with QPCs midway through the channel.  The precise QPC geometry can be seen in Fig.~1 of the main text.

\section*{S3. Further insights from Fig. 2d}

We examine several locations in Fig.~2d in closer detail, to see how features in this image result from temperature-induced screening.  The drain is on the right (source on the left), so heating the drain causes electrons to flow to the left (right) when $dt/d\varepsilon$ is positive (negative).  \textbf{We define positive contributions to $\tilde S$ to correspond to more current flow to the left when $T_\mathrm{d}$ increases.}  Schematics indicate how the bottom of 1D subbands vary with position along the QPC. Energy is highest at the saddle point (in the middle of the QPC) and falls off towards source or drain leads, to the left or right respectively.

\hspace{-0.2in}\includegraphics[width=0.96\textwidth]{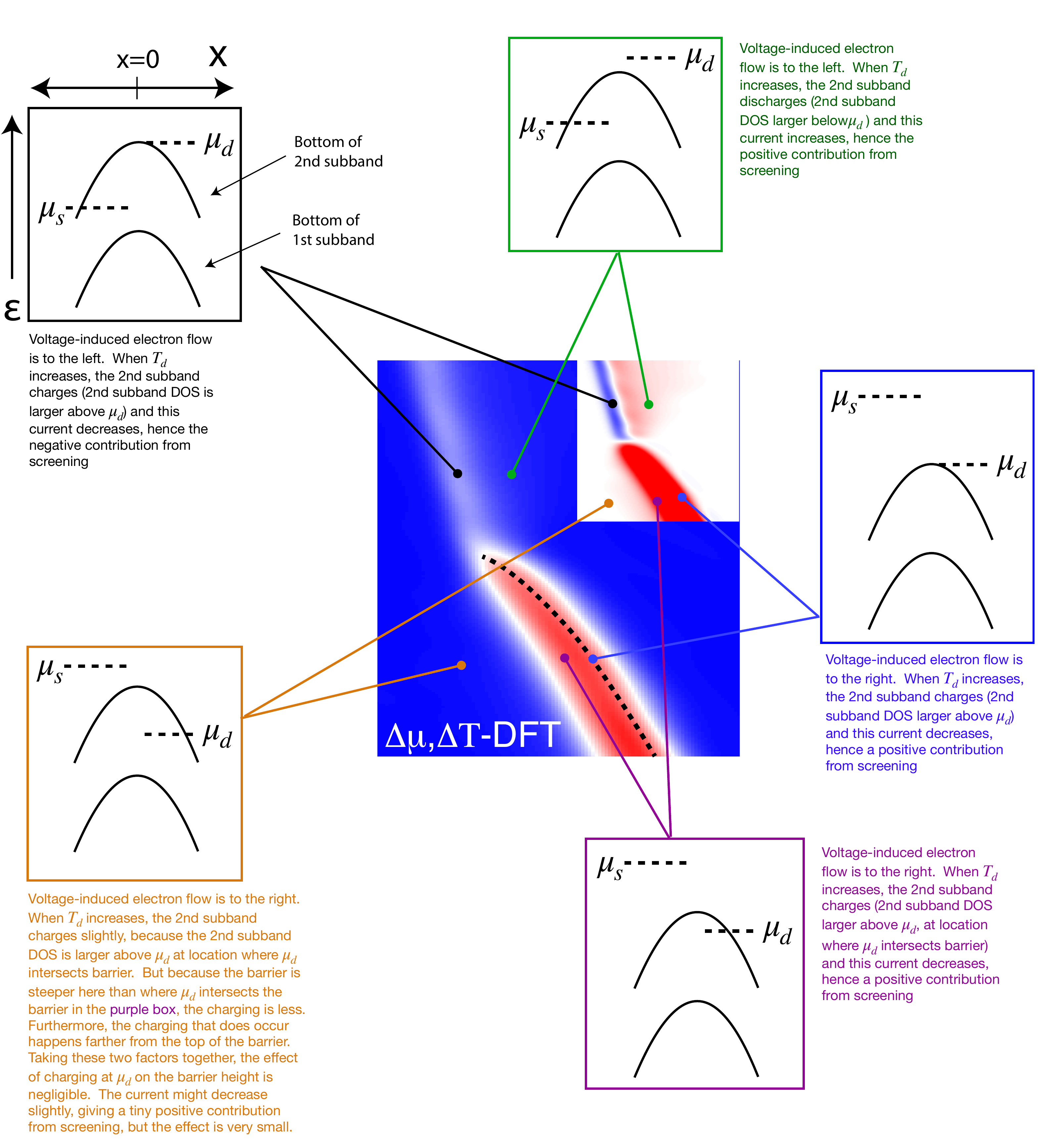}

\clearpage
\section*{S4. Further discussion about point X$_2$}

Here we present evidence that the extra leg in the thermopower at point X$_2$ in Fig.~3b of the main article is due to presence of a localized state near the top of the QPC barrier which makes the potential there very sensitive to heating induced charging of the localized state at the drain chemical potential. 

\begin{figure}[h]
\includegraphics[width=0.8\textwidth]{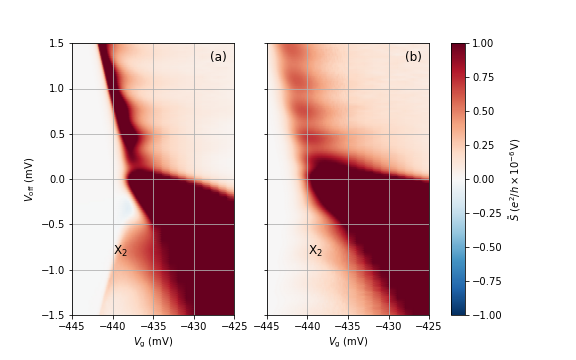}
\caption{$\Delta\mu$,$\Delta T$-DFT thermopower simulations (a) with (same as Fig.~3b in the main article) and (b) without exchange-correlation terms. X$_2$ marks the points at $V_\mathrm{g}=-439\,$mV and $V_\textrm{off}=-0.8\,$mV. \label{xcnoxc}}
\end{figure}

\begin{figure}[h]
\includegraphics[width=1\textwidth]{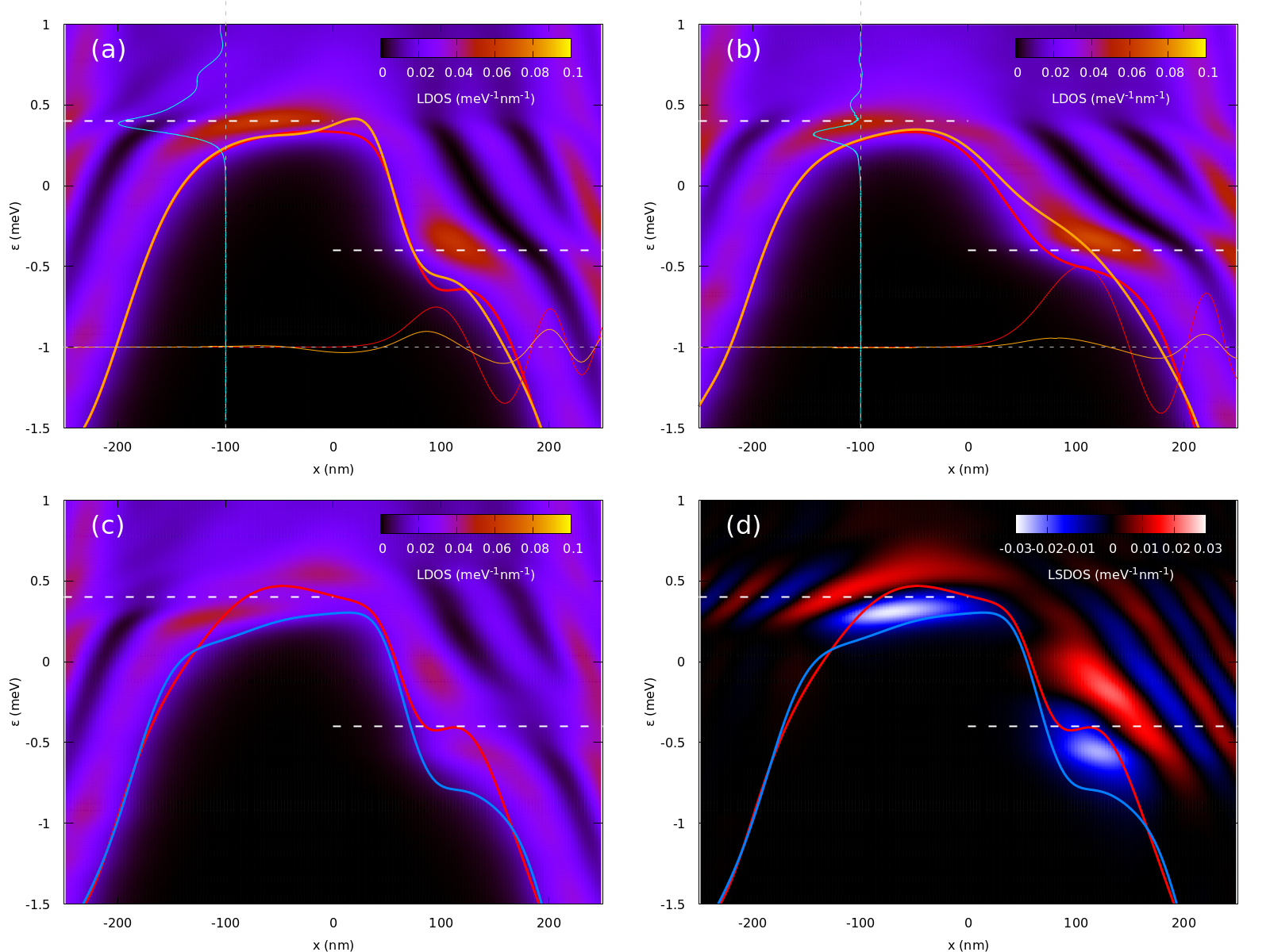}
\caption{(a) LDOS at point X$_2$ in Fig.~\ref{xcnoxc}a. The thick red and orange lines are the first subband QPC potentials at $T_\mathrm{s}=T_\mathrm{d}=100\,$mK and at $T_\mathrm{s}=100\,$mK, $T_\mathrm{d}=150\,$mK (the difference to the red line is multiplied by 100), respectively. The long-dashed white lines are the source and drain chemical potentials. The thin red and orange lines show the charging (in $10^{-4}\,$nm$^{-1}$, offset vertically by -1) induced by heating the drain in $\Delta\mu$-DFT and $\Delta\mu$,$\Delta T$-DFT simulations, respectively. The cyan line shows the drop of the transmission function due to the temperature induced screening (multiplied by 500 and offset horizontally by -100). (b) The same at point X$_2$ in Fig.~\ref{xcnoxc}b. (c)~LDOS and first subband  spin-up and spin-down QPC potentials (red and blue lines, respectively) in the LSDA simulation with exchange-correlations terms. (d) Local spin density of states (LSDOS) corresponding to panel (c). \label{ldos}}
\end{figure}

Note in Fig.~\ref{xcnoxc}b that this extra leg is not present in the $\Delta\mu$,$\Delta T$-DFT simulation if the exchange-correlation term is not taken into account, i.e. in the Hartree aproximation. The sensitivity of the potential at the top of the barrier to charging due to heating of the drain is presented in Figs.~\ref{ldos}a,b for simulations with and without exchange-correlation term, respectively. Starting with both source and drain at the same temperature (the potentials for this case are plotted with thick red lines), we heat the drain which causes charging (thin red lines) near the point where the drain chemical potential intersects the QPC potential. Electrons in vicinity screen this additional charge, resulting in a different QPC barrier (thick orange lines) and a different distributions of charge (thin orange lines). Because the potential barriers grow higher, the transmission functions become smaller (cyan lines), leading to a positive thermopower, as explained in the main article. 

Notice that the potential at the top of the barrier is much more sensitive to heat induced charging when the exchange-correlation term is taken into account. To see why this is so, consider Figs.~\ref{ldos}c,d showing the local total and spin densities of states, respectively, calculated at point X$_2$ using the spin density functional theory (SDFT). Contrary to the unpolarized DFT calculations used in the main article and in Fig.~\ref{ldos}a, SDFT allows spin polarization to take place which is favorable in the regions where the density is low, i.e. near the points where a chemical potential intersects the QPC potential. To the left of point X$_2$, i.e., at more negative gate voltages, the QPC barrier rises above both chemical potentials and there are two localized states, one on each side of the barrier. At point X$_2$, the barrier has lowered sufficiently that the localized state pinned to the source chemical potential has spread over the top of the barrier (Figs.~\ref{ldos}c,d). Increasing the gate voltage further, the barrier drops far below the source chemical potential and this localized state is replaced by a smooth unpolarized charge density. The spreading of the localized state across the top of the barrier and then its eventual disappearance occurs over a narrow interval of gate voltages. Therefore, near X$_2$ the state of the QPC is very sensitive to additional charge appearing in vicinity. The formation of localized states at low densities can be traced to the exchange interaction. In the Hartree approximation, the evolution in this gate voltage interval is much smoother and, therefore, such sensitivity is not observed.

\clearpage

\par

\bibliography{2010QPCbib}